\documentclass[a4paper,11pt]{article}
\pdfoutput=1 

\usepackage{jcappub} 
\usepackage{pgfplots}
\usepackage{graphics}      
\usepackage{graphicx}      
\usepackage{longtable}     
\usepackage{url}           
\usepackage{bm}            
\usepackage{gensymb}
\usepackage{comment}	   
\usepackage[english]{babel}
\usepackage{physics}
\usepackage{multirow}
\usepackage{amsmath, amsthm, amsfonts, amssymb, mathtools, calrsfs, tensor, tikz-cd}
\usepackage{cases}
\pgfplotsset{width=7cm,compat=1.3}
\DeclareMathOperator{\Erf}{Erf}
\begin{document}
\raggedbottom

\title{Macro detection using fluorescence detectors}
\author[a]{Jagjit Singh Sidhu}
\author[a]{Roshan Mammen Abraham}
\author[a]{Corbin Covault}
\author[a]{Glenn Starkman}
\affiliation[a]{Physics Department/CERCA/ISO Case Western Reserve University  \\
	Cleveland, Ohio 44106-7079, USA}
\emailAdd{jxs1325@case.edu}
\emailAdd{rma93@case.edu}
\emailAdd{cec8@case.edu}
\emailAdd{gds6@case.edu}
\abstract{Macroscopic dark matter (aka {\em macros}) constitutes a broad class of alternatives to particulate dark matter. We calculate the luminosity produced by the passage of a single macro as a function of its physical cross section. A general detection scheme is developed for measuring the fluorescence caused by a passing macro in the atmosphere that is applicable to any ground based or space based Fluorescence Detecting (FD) telescopes. In particular, we employ this scheme to constrain the parameter space ($\sigma_{x} \mbox{ vs} \mbox{ M}_{x}$) of macros than can be probed by the 
	Pierre Auger Observatory and by the  
	Extreme Universe Space Observatory onboard the Japanese Experiment Module 
	(JEM-EUSO). It is of particular significance that both detectors are sensitive to macros
	of nuclear density, 
	since most candidates that have been explored (excepting primordial black holes) 
	are expected to be of approximately nuclear density.}
\keywords{Macroscopic Dark Matter, Dark Matter, Fluorescence Detectors, Pierre Auger, JEM-EUSO}
\arxivnumber{}

	\maketitle

\section{Introduction}
If General Relativity is correct, then dark matter constitutes most of the mass density of the Galaxy. While dark matter is widely thought to exist (although see \cite{a}) we have yet to detect it except gravitationally. 

The most widely considered and searched for candidates are new particles not found in the Standard Model of particle physics, such as the generic class of Weakly Interacting Massive Particles (WIMPs) (especially the Lightest Supersymmetric Particle) and axions. Recently, renewed attention has been paid to primordial black holes and to macroscopic composite objects, aka macros, especially those of approximately nuclear density.
The theoretical motivation for this 
stems originally from the work of Witten \cite{b}, 
and later, more carefully Lynn, Nelson and Tetradis \cite{c},
Macroscopic objects made of baryons may be stable with sufficient strangeness, 
and may have formed before nucleosynthesis \cite{b,c}, 
thus evading the principal constraint on baryonic dark matter. 
One appeal of such a dark matter candidate is that there would be no need to invoke the existence of new particles to explain the observed discrepancy between gravitational masses and luminous masses in galaxies.  Numerous beyond-the-Standard-Model macro candidates have also been suggested (e.g., 
\cite{d}).

Recently one of us (GDS), along with colleagues, presented a comprehensive assessment of limits on such macros as a function of their mass and cross-section \cite{e}, identifying specific windows in that parameter space that were as yet unprobed. (We later refined those  \cite{f}.)

Taking macros to interact with our detectors with their geometric
cross section, the expected number of macro events detected by an
observatory/detector with effective area $A_{ef}$ that operates continuously over
an observing time $t_{obs}$ is given by
\begin{equation} \label{eventrate}
	\begin{aligned}
		N_{events}&=\dfrac{\rho_{_{DM}}}{M_{x}}A_{ef} t_{obs} v_{x} \\
		&=5.5\left(\frac{kg}{M_{x}}\right) \left(\frac{A_{ef}}{1000\ km^2}\right)
		\left(\frac{t_{obs}}{yr}\right)
	\end{aligned}
\end{equation}
where $\rho_{_{DM}}$ is the local dark matter density $7\times10^{-25}\,$g cm$^{-3}$
\cite{e}, and $M_x$ is the mass of the macro.  For the
purposes of this paper, we assume macros possess a Maxwellian distribution of speeds given by
\begin{equation}
f(v_x)_{MB} = \left( \frac{1}{\pi v_{vir}^2}\right)^{\frac{3}{2}}4\pi v_x^2 e^{-\left(\frac{v_x}{v_{vir}}\right)^2}, \label{eq:maxwellian}
\end{equation}
where $v_{vir} \approx 250\,$km s$^{-1}$.  This distribution is slightly modified by the motion of the Earth
as described in detail in Section~\ref{sec:fofvx}. 
The cumulative distribution function is then 
obtained by integrating the probability 
distribution function up to the desired value of 
$v_x$. This allows us to determine the maximum $M_x$ we can probe as a function of $v_x$.
 


With a minimum allowed macro mass of $55\,$g (inferred from
mica\cite{e} that has been ``exposed'' to the bombardment of macros for tens
of millions of years) the number density, and hence flux, of macros is
quite small.  Thus, any plan to detect macros on human time scales
({\it e.g.}, years) requires a target of very large area.

In this work, we explore the possibility that fluorescence detectors designed to detect ultra-high
energy cosmic rays might be simply modified and effectively used to detect the nitrogen fluorescence caused by a macro's passage through the atmosphere.
Through elastic scattering, the macro would deposit enough 
energy to dissociate the molecules and ionize or excite the atoms. 
This results in the formation of a plasma.

\begin{figure}
   \begin{center}
     \includegraphics[width=6.00in]{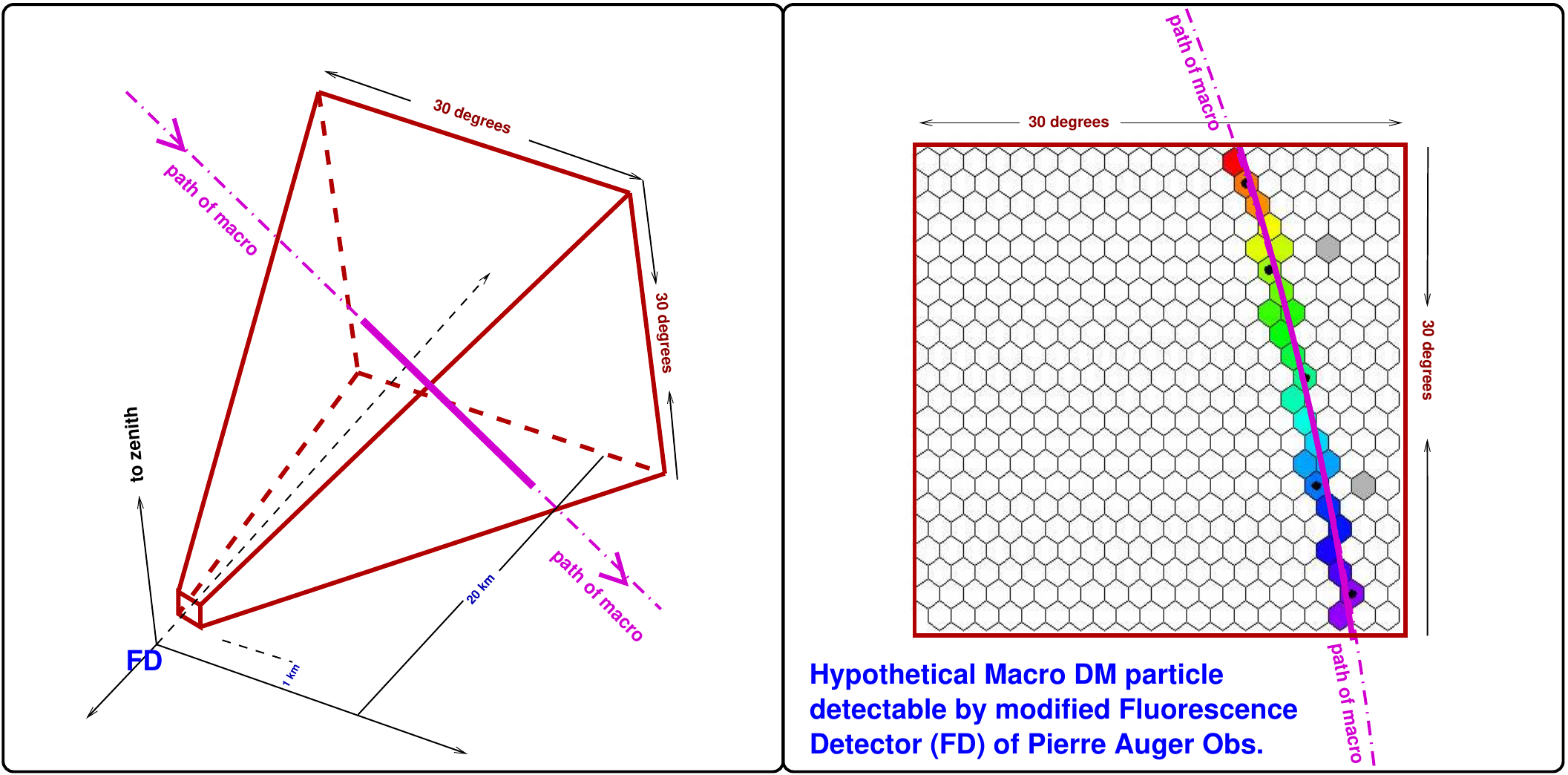}
        \caption{Conceptual diagram delineating the plausible
          detection of a macro dark matter particle by a modified
          version of a Fluorescence Detector (FD) such as that of the
          Pierre Auger Observatory.  {\bf Left:} Perspective view of
          the truncated pyramid corresponding to the $30^{\circ}
          \times 30^{\circ}$ field-of-view of a single camera (`eye')
          with a fiducial detection volume that ranges from 1~km to
          20~km distance from the Auger FD.  In this example, the
          macro particle penetrates the fiducial region at a
          near-vertical angle generating a perfectly straight ``pencil
          beam'' of ionized plasma. Note that in contrast to
          relativistic cosmic rays, a macro particle is moving much
          slower and will not generate an air shower.  {\bf Right:}
          Example `event display' for a modified Auger FD showing the
          predicted appearance of a macro signal detected within the
          field-of-view (the linear path of the macro appears as a
          slightly bent line due to curvature of the FD camera focal
          plane).  The three-dimensional path of the macro can be
          fully reconstructed using pixel locations and timing.  A macro
          event is clearly distinguished from cosmic rays and other
          atmospheric phenomena by it's velocity (about 250~km/s),
          perfectly straight path, and an ionization signal
          proportional to the atmospheric density.}
\end{center}        
\label{fig:roughidea}
\end{figure}

Figure~\ref{fig:roughidea} demonstrates the concept of
  detection of a macro dark matter in the atmosphere by, for example,
  a modified version of a single Fluorescence Detector (FD) telescope
  of the Pierre Auger Observatory.  In this example, the macro
  particle penetrates the atmosphere generating a narrow column of
  ionization within the field of view of the FD.

We show below that the size 
of the macro will determine the size of the resulting plasma. For 
large enough values of the macro cross-section $\sigma_x$, 
we find that the plasma becomes optically thick to photons and radiates as a 
blackbody. We analyze the optically thick and thin mechanisms separately.

As heat diffuses out of this region surrounding the macro trajectory, 
the ions recombine to release photons that can be detected using 
fluorescence detectors. However, the macro passage through the 
atmosphere does not create an air shower as when a high energy cosmic ray is detected.
\color{black}We find that for the Pierre Auger Observatory (Auger) and the proposed Extreme Universe Space Observatory onboard the Japanese Experiment Module (JEM-EUSO), we can probe masses up to 1.6$\times 10^{4}\,$g and 5.5$\times 10^{6}\,$g respectively for an observation period of 1 year, thus providing significant improvements over the current $55\,$g limit.


\section{Fluorescence Signal}

The cross-sectional area $\sigma_x$ of a macro of mass $M_x$ and density $\rho_x$
is 
\begin{eqnarray}
\sigma_x = 2.4\times10^{-10}\,{\mathrm {cm}}^2 
	\left(\frac{M_x}{g}\right)^{2/3}\left(\frac{\rho_{nuclear}}{\rho_x}\right)^{2/3}\,,
\end{eqnarray}
where we take $\rho_{nuclear}=3.6\times10^{14}\,$g cm$^{-3}$.
In this manuscript, we are interested in probing densities above $10^6\,$g cm$^{-3}$ up to nuclear densities,
with a particular emphasis on the latter.
A macro of mass greater than $55\,$g and of a density of likely interest 
will therefore be much larger than the separation of molecules in the atmosphere, 
	$\sim$ few $\times10^{-7}\,$cm.
We also reasonably assume that the internal density of the macro is much larger than that of air, and that its velocity
will not be significantly altered as a result of collisions with individual air molecules.

During its passage through the atmosphere, the macro will dissociate
molecules, and ionize and excite atoms both by direct impact, more
importantly through subsequent secondary effects of those impacts.
With $v_x= 250\,$km s$^{-1}$, after the impact of a macro with a single
nitrogen molecule, that molecule will be dissociated, and the nitrogen
atoms will rebound with velocities $v\lesssim 500\,$km s$^{-1}$.  The
kinetic energy of the rebounding atoms will result in secondary
collisions with other air molecules.  Since many secondary collisions
are required before the energy is thermalized, these collisions will
dominate the net ionization and excitation.

Following the work of Cyncynates et al. \cite{g}, 
we propagate the initial energy deposition by the macro,
which we approximate as a delta function along a straight trajectory, outward
radially away from that trajectory according to the heat equation.
Ignoring radiative cooling, the temperature field after some time $t$ is 
\begin{equation}\label{tempfield}
	T(r,t) = 
	\frac{\sigma_{x} v_x^2}{4\pi \alpha c_p}\frac{e^{-\frac{r^2}{4t\alpha}}}{t},
\end{equation}
where  $\alpha \approx 10^{-4}\,$m$^2\,$s$^{-1}$exp$(D/10$km$)$ 
	is the thermal diffusivity of the air,
and $c_p \approx 25$ kJ kg$^{-1}$ K$^{-1}$ 
	is the specific heat of the air \cite{h}.
(The specific heat varies around a mean of $\sim25\,$kJ kg$^{-1}\,$K$^{-1}$ 
	for temperatures between $10^4\,$K and $10^5\,$K,
and the corresponding thermal diffusivities vary around $0.08\,$m$^2\,$s$^{-1}$.)

We invert \eqref{tempfield} to obtain $\pi r_I(t)^2$, 
the area at time $t$ that reaches a particular state of ionization I 
characterized by the appropriate ionization temperature $T_I$,
by setting $T(r,t) = T_I$. 
This area is given by
\begin{equation}\label{ionizationcrosssection}
\pi r_I(t)^2 = 4\pi\alpha t\log\left(\frac{\sigma_{x} v_x^2}{4\pi \alpha t c_p T_I}\right) .
\end{equation}
After the macro passes, 
the ionized area starts at $0$ at $t=0$, increases to a maximum,
then falls back to $0$ at

\begin{equation}\label{coolingtime}
t_{I0}=\frac{\sigma_x v_x^2}{4\pi \alpha c_p T_I} \approx 20\,s\left(\frac{\sigma_x}{cm^2}\right)\left(\frac{v_x}{250kms^{-1}}\right)^2e^{-\frac{D}{10 km}}.
\end{equation} 

$t_{cool}$ is defined as the time for the 
temperature everywhere to fall below 
	$T_I = 10^4\,$K\cite{i} corresponding to the ionization temperature for NI. 
At this point nearly all the electrons will have recombined 
and hence $\pi r_I(t)^2 = 0$. In other words, $t_{cool} = t_{I0}$. 

\begin{figure}
 \centering

        \includegraphics[width=4.00in]{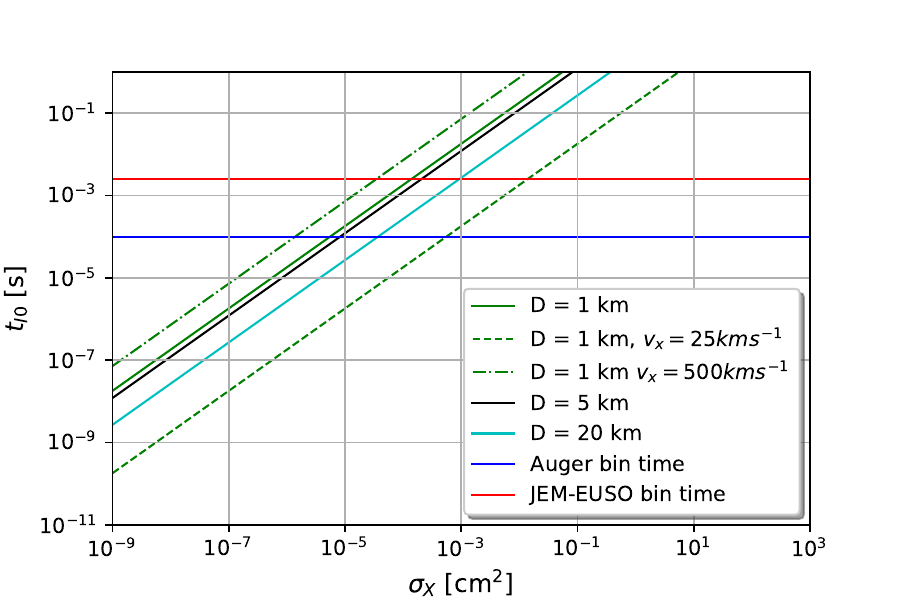}
        \caption{Cooling time as a function of macro cross section $\sigma_x$. The two horizontal lines represent the proposed bin times for Auger and JEM-EUSO to be able to detect macros. }
        \label{fig:bintime}

 \end{figure}

The cooling time (equation 
\eqref{coolingtime}) thus depends on the 
cross section $\sigma_x$ and the speed $v_x$ of the 
incoming macro. Typically, most incoming 
macros will approach at \mbox{$v_x \sim 250\,$km s$^{-1}$.}
Thus the cooling time is plotted taking $v_x = 
250\,$km~s$^{-1}$ and $D = 1$~km, $5$~km and $20$~km respectively 
 in Figure 
\ref{fig:bintime}, along with the proposed bin 
times for Auger and JEM-EUSO to be able to detect 
macros. In particular, if the bin time of an FD 
is less than the cooling time, $t_{cool}$, the 
macro will cause fluorescence over multiple time 
bins. We proceed to calculate the number of 
photons produced during this interval.

The formation of a plasma with a definite 
temperature field will not hold for macros of a 
small $\sigma_x$ which would be unable to impart 
sufficient energy to a large number of nitrogen 
molecules.  The (internal) energy fluctuates
randomly about the mean value. This statement can be quantified by the thermal fluctuations, which are given by
\begin{eqnarray}
\left(\frac{\langle (E-\langle E\rangle)^2 \rangle}{\langle E \rangle^2}\right) \sim N^{-\frac{1}{2}}
\end{eqnarray}
The quantity N is determined by the causal volume 
of the resulting plasma, i.e. $N = \sigma_x n_a 
L$, where $L = c_s t_{I0}$ and $c_s \approx 
300\,$m s$^{-1}$ is the speed 
of the sound in air.
By requiring that the thermal fluctuations do not 
exceed $10\%$, i.e. $N \geq 100$, we find 
$\sigma_x \gtrapprox 3\times 10^{-12}$ cm$^2$ 
will produce a plasma. This sets a lower bound on 
the sensitivity of an FD to detect macros. 

All subsequent analysis will 
be to determine if an FD can reach this lower 
limit. Each FD produces an effective volume that 
relates the height, D, away from the pixels to 
the maximum mass, $M_x$ that FD can probe at that 
particular height. In other words, $M_x(D)$ for a 
given integration time. Such a relationship 
allows us to relate $\sigma_x$ to $M_x$ to 
quantify a lower bound that can be plotted on 
\ref{fig:exclusion}. This analysis is done for two particular FDs in section 4.

An important distinction needs to be made regarding the opacity of the resulting plasma along the macro trajectory. For large enough $r_I$, the plasma becomes optically thick.  
For the temperatures of interest ($\geq10^4\,$K), 
the scattering length of air is 
a few meters \cite{j}. 

To obtain the value of $\sigma_x$  
demarcating the transition between an optically thick and thin plasma, 
we first find the maximum area that achieves each specific degree of ionization
\begin{equation}\label{max}
\pi r_{I,max}^2 = 4t_{I,max}\alpha\pi =\frac{\sigma_x v_x^2}{c_p T_I e}\,,
\end{equation}
reached at 
$t_{I,max} = t_{I0}/e$.  
The $T_I$ of interest corresponds to that of NII. 

We compare $r_{I,max}$ and the scattering length 
$\lambda_{sc}$, which is dependent on the number 
density of the atoms and hence the height above 
the ground
\begin{equation}\label{scatlen}
\lambda_{sc} = \lambda_{sc,0}e^{\frac{D}{10 \,km}}
\end{equation}
where $\lambda_{sc,0}$ is the scattering length 
for the number density of the atoms at ground 
level.
Using \eqref{scatlen} in \eqref{max}, we find
that only large macros with cross sections at the 
upper limits of the unprobed parameter space
could plausibly produce a large enough column of 
plasma that would emit as a blackbody in the 
optically thick regime, i.e macros that satisfy
\begin{equation}\label{transition}
\left(\frac{\sigma_x}{cm^2}\right)\left(\frac{v_x}{250km\,s^{-1}}\right)^2 \geq e^{\frac{D}{5 \,km}}22000
\end{equation}
For completeness, we present that case separately from 
the optically thin emission mechanism below.

Another important consideration is the relationship between 
the recombination timescales of the electrons through three-body recombination, radiative recombination and spontaneous emission.
To the best of our knowledge, the rates that exist in the literature are from very low electron number densities that are not applicable to our analysis. 
The NII rates, however, are available for reasonably high electron number densities. Using the rates from 
\cite{j1} (and also see references therein)
we find that approximately $10^{-6}$ electrons 
will recombine radiatively. In \cite{j1} these 
rates were calculated for both an optically thick 
and thin plasma. The optically thin calculation 
depends on both the electron temperature and 
number density(see Figures 8 and 9 of \cite{j1}). 
The rates thus depend on the atmospheric depth 
through the number density of electrons. The 
discrepancies in these quantities with 
experiments are approximately a factor of 2 
\cite{j1} and thus we expect 
corrections to be of order $\sim 1$ to the SNR values calculated below.
We account for this below when we calculate the number of photons emitted 
through a factor 
\begin{equation}
\beta_{1} = \frac{\alpha_{rad}}{k_{three}n_e} = 10^{-6}
\end{equation}that represents the fraction of electrons  
that recombine radiatively for the first ionization state.
We have used the three-body recombination coefficient for $n_e = 10^{22}\,$m$^{-3}$. This is an underestimate of the true value of $\beta_I$ because the coefficient decreases with increasing $n_e$. However, reference \cite{j1} only obtained the three-body recombination coefficient up to $n_e = 10^{22}\,$m$^{-3}$.

For  photons free-streaming from the optically thin plasma, 
	{\it i.e.}  $r_{I,max} < \lambda_{sc}$, 
we must also compare the ``recombination time'' of the plasma $t_{rec}$
to its ``cooling time'' $t_{cool}$.
If $t_{rec} \leq t_{cool}$, 
each electron can contribute multiple photons 
	through multiple ionizations and recombinations. 

The radiative recombination rate coefficient 
	is approximately constant 
	across the temperature range of interest \cite{j1}, 
$\alpha_{rad} \approx 4\times 10^{-19}\,$m$^3\,$s$^{-1}$. 
Thus  
\begin{equation}\label{trec}
t_{rec} \equiv  \frac{1}{n_e \cdot \alpha_{rad}} \simeq 3\times 10^{-8} s \ e^{\frac{D}{10 \, km}},
\end{equation}
with $n_e = n_{e,0} exp(-D/10$km$)$ and $n_{e,0} \approx 10^{26}\,$m$^{-3}$ is the electron number density at ground level.

Comparing $t_{rec}$ with $t_{cool}$, we find
\begin{equation}
\left(\frac{\sigma_x}{cm^2}\right) \geq 10^{-9}\, e^{\frac{D}{5\, km}}\left(\frac{250\, km\,s^{-1}}{v_x}\right)^2.
\end{equation}
For the values of $\sigma_x$, $v_x$ and D of 
interest, we 
find that macros in the parameter space we seek 
to probe would produce a plasma where each 
electron will recombine multiple times. 

To account for this, we multiply the expression 
for the number of photons produced per unit 
length $\frac{dN}{dl}\gamma$ by a quantity 
$N_{mult}= \min\left(1,\frac{t_{cool}}{t_{rec}}\right)$.

To find the number of photons that reach a 
detector pixel, 
we first estimate the electron number density in 
the recombining plasma, 
by using \eqref{ionizationcrosssection} for $t_{I0}/e \leq t \leq t_{I0}$.
We can then find the number of photons emitted during recombination. 
We multiply  by the transverse distance L $=$ D$\,$sin$\theta$(we justify this choice for L in section 5)
traveled by the macro across the field of view of a pixel
(where D is the height of the macro path above the detector 
and $\theta$ is the angle viewed by the pixel)
and by the ratio of the  detector area to the area of a sphere of radius D,
\begin{equation}\label{receivedatdetector}
N_{\gamma} = \dv{N}{l} \left(\frac{A_{detector}}{4\pi D^2}\right) L 
	= \dv{N}{l}\left(\frac{ A_{detector} \sin\theta}{4\pi D}\right).
\end{equation}

For the free streaming case, 
photon production happens on a timescale $t_{I0}$. 
A Fluorescence Detector (FD) telescope 
integrates over small intervals, 
known as the bin time, $\tau_{bin}$. 
For the $\tau_{bin}$ we consider below,
$t_{I0} < \tau_{bin}$ for all $\sigma_{x}$. 
Consequently the number of photons emitted 
per unit length along the macro path trajectory in one bin time is

\begin{equation}\label{freestreaming}
	\begin{aligned}
	\dv{N}{l} &= 2 N_{mult} \eta \pi  \int_{\frac{t_{I0}}{e}}^{t_{I0}}\int_{0}^{r_I(t)} 
		\beta_1 n_{e1} n_a \alpha_{rad} r dr dt \\ 
		&\approx 2 N_{mult} \times10^{-8}s^2\alpha \eta  
			n_a^2 \alpha_{rad}\left(\frac{\sigma_x}{cm^2}\right)^2\left(\frac{v_x}{250 \, km s^{-1}}\right)^4
		\\ \\
	\end{aligned}
\end{equation}


We have introduced in \eqref{freestreaming} the quantity $\eta$, 
	which is the probability that transitions in a nitrogen plasma produce a photon in the $350 - 400\,$nm detection range. 
This is important as the detectors considered below are sensitive
only within this waveband. 
This value was obtained from the ratio of Einstein coefficients 
	of transitions that resulted in wavelengths within our range 
	to the Einstein coefficients of all transitions 
	(using tabulated data from \cite{m})
\begin{equation}
\eta = \frac{\sum\nolimits_{350-400nm} A_{ki}}{\sum\nolimits_{all} A_{ki}},
\end{equation}
where $A_{ki}$ is the Einstein coefficient for a transition from a state i to a state k.  
We find $\eta \approx 2\times10^{-3}$.

The quantity $N_{mult}$ is given as 
\begin{equation}\label{nmult}
N_{mult}= \frac{\rho_0 \sigma_x v_x^2 \alpha_{rad} n_{e0} e^{-\frac{D}{5}}}{4\pi K T_1} = 6\times 10^8 \left(\frac{\sigma_x}{cm^2}\right)
\left(\frac{v_x}{250\,km\,s^{-1}}\right) e^{-\frac{D}{5\,km}}
\end{equation}
Evaluating \eqref{receivedatdetector} using 
\eqref{freestreaming} and \eqref{nmult}, 
the number of photons received in a detector 
pixel in one bin time
in the case of the optically thin plasma is
\begin{equation}\label{freestreamingevaluated}
		N_{\gamma}^{thin} = 3 \times10^{27} \min\!\left(\frac{\tau_{bin}}{t_{I0}},1\right)
		\frac{A_{detector} \sin\theta }{D~km} 
			\left(\frac{\sigma_x}{cm^2}\right)^3\left(\frac{v_x}{250\,km\,s^{-1}}\right)^6e^{-\frac{2D}{5\,km}}\,
\end{equation}
where $A_{detector}$ and D have units of km$^2$(as will be the case throughout this analysis).
The factor $\min\!\left(\frac{\tau_{bin}}{t_{I0}},1\right)$ accounts
approximately for the case where $\tau_{bin} \leq t_{I0}$, with the
precise value of $N_\gamma^{thin}$ depending on when exactly the light
from the macro path enters the pixel. 

Since we ignored radiative cooling when solving the heat equation, 
we calculate the fraction of the energy deposited by the macro that we have attributed to photon emission to ensure that we do not violate energy conservation 
\begin{equation}
	\begin{aligned}
		\frac{E_{freestream}}{E_{initial}}&=\frac{\mbox{N}_\gamma^{thin}\overline{E}}{\frac{1}{2}\rho_{atm}(v_x)^2\sigma_{x}\mbox{L}}\\
		&\approx 2\times 10^5 \left(\frac{\sigma_x}{cm^2}\right)^2\left(\frac{v_x}{250\,km\,s^{-1}}\right)^4 e^{-\frac{2D}{5\,km}},
	\end{aligned}
\end{equation}
For $\left(\frac{\sigma_x}{cm^2}\right)^2\left(\frac{v_x}{250\,km\,s^{-1}}\right)^4 \gtrapprox \frac{1}{2\times 10^5}$,this implies a violation of energy conservation that may be traced back to the failure to include radiative losses in the solution \eqref{tempfield} of the heat equation. 
(As appropriate in the context of \cite{g}.)
The value of $N_{\gamma}^{thin}$ from \eqref{freestreamingevaluated} 
should not be trusted 
but would instead be expected to 
approximately saturate at the level
\begin{equation}\label{fsfinal}
	N_{\gamma}^{thin} = 3 \times10^{19}
		\frac{A_{detector} \sin\theta }{D~km} 
			\,\min \left(10^{8} \left(\frac{\sigma_x}{cm^2}\right)^3\left(\frac{v_x}{250\,km\,s^{-1}}\right)^6e^{-\frac{2D}{5\,km}}, 1 \right) \min\!\left(\frac{\tau_{bin}}{t_{I0}},1\right)
\end{equation}
 
We now proceed to examine the blackbody emission case. 
We integrate the Planck spectrum 
over the wavelengths of interest 
to find the number of photons emitted by the plasma
per unit length along the macro trajectory 
\begin{equation}\label{bb}
	\begin{aligned}
	\dv{N}{l} &= \int_{0}^{t_{cool}}\int_{\nu_{350nm}}^{\nu_{400nm}} 
		\frac{4\pi 2\pi r{\tiny _{10000K}}(t)}{h\nu} B(\nu, T)d\nu dt\\ 
	&= 3.0 \times 10^{27}\frac{1}{km} \left(\frac{\sigma_x}{cm^2}\right)^{3/2}
	\left(\frac{v_x}{250\,km\,s^{-1}}\right)^3
	\end{aligned}
\end{equation}
From \eqref{receivedatdetector} and \eqref{bb}, 
we get $N_{\gamma}^{thick}$ just as in the free-streaming case

\begin{equation}\label{bbevaluated1}
N_{\gamma}^{thick} = 2\times10^{26} \min\!\left(\frac{\tau_{bin}}{t_{I0}},1\right)
	\frac{A_{detector} \sin\theta}{D~km} 
	\left(\frac{\sigma_x}{cm^2}\right)^{3/2}\
	\left(\frac{v_x}{250\,km\,s^{-1}}\right)^3
\end{equation}
\color{black}

We similarly calculate the energy fraction that is emitted as photons for the blackbody scenario by
\begin{equation}\label{blah}
	\begin{aligned}
		\frac{E_{bb}}{E_{initial}}
        \approx 0.4 \left(\frac{\sigma_x}{cm^2}\right)^{\frac{1}{2}}\left(\frac{v_x}{250kms^{-1}}\right)
\end{aligned}
\end{equation}

For $\left(\frac{\sigma_x}{cm^2}\right)^{\frac{1}{2}}\left(\frac{v_x}{250\,km\,s^{-1}}\right) \gtrapprox 1$, this implies a violation of energy conservation as in the case of the optically thin solution. Additionally, the value of the cooling times, $t_{I0}$ are much longer than the bin time that the factor $\min\!\left(\frac{\tau_{bin}}{t_{I0}},1\right)$ is always given by $\frac{\tau_{bin}}{t_{I0}}$. 
The value of $N_{\gamma}^{thick}$ from \eqref{bbevaluated1} 
would be expected to 
saturate at 

\begin{equation}\label{bbevaluated}
N_{\gamma}^{thick} = 3\times10^{24} \frac{A_{detector} \sin\theta}{D~km}
	\times
	\frac{\tau_{bin}}{t_{I0}}
\end{equation}
where we have taken $\left(\frac{\sigma_x}{cm^2}\right)^{\frac{1}{2}}\left(\frac{v_x}{250\,km\,s^{-1}}\right) \approx 1$ from \eqref{blah}. The blackbody regime exists entirely above the $\sigma_x v_x^2$ cutoff from \eqref{blah}.

 \begin{figure}
 \centering
        \includegraphics[width=4in]{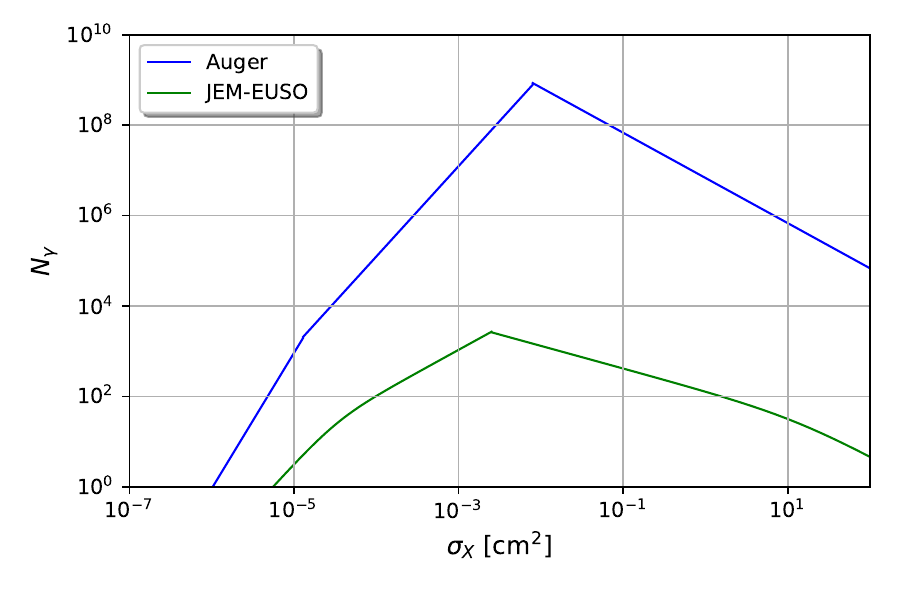}
        \caption{The expected number of photons received at the detector as a function of $\sigma_x$ for Auger(green) taking D$=10\,$km and JEM-EUSO(black) taking D$=400\,$km as predicted in section 4.1. The choice of D $= 10$ km for Auger is explained in section 4.1.}
        \label{fig:sigmadependence}
 \end{figure}
\eqref{fsfinal} and \eqref{bbevaluated} are multi-valued depending on the value of $\sigma_x$ and $v_x$. For large enough or sufficiently quick macros, a pixel can be lit up for multiple time bins. In addition to that, the amount of photons produced levels off for sufficiently large or high speed macros. 
For both detectors considered in this work, the transition to pixels being lit up for multiple time bins occurs prior to the leveling off. 
In Figure \ref{fig:sigmadependence}, we plot \eqref{freestreamingevaluated} and \eqref{bbevaluated} for both Auger(with a characteristic D$ = 10\,$km) and JEM-EUSO across the entire range of $\sigma_x$ we hope to probe. We provide detector specific quantities used in this calculation in \eqref{freestreamingevaluated} and \eqref{bbevaluated} in section 4 below.

Since we have solved for the number of photons produced in the optically thick and thin cases separately, we find that the solutions do not smoothly transition \eqref{transition}. Thus, we interpolated the optically thin solution at the transition points. We expect that after 3 scattering lengths that thermalization will occur and the optically thick regime is thus set at $9 \sigma_{x,FS}$, where $\sigma_{x,FS}$ is the transition cross section for a fixed $v_x$.

In the following section, we discuss the signal-to-noise ratio.

\section{Detection Signal-to-Noise}
In the previous section, 
we calculated the number of photons $N_\gamma$
produced by the macro in the detector waveband
that reach a detector pixel. 
To ensure an FD telescope can detect the macro
requires that the signal produced by those photons 
exceed the noise due to the background, i.e.
we require that the signal-to-noise ratio (SNR)
\begin{equation}\label{SNR}
\mbox{SNR}=\dfrac{Q \ F N_{\gamma}}{N_{noise}} \,,
\end{equation}
exceed some threshold value.

As we see in \eqref{SNR}, the signal consists of the the number of
photons incident on the detector $N_\gamma$ multiplied by the
photomultiplier tube (PMT) quantum efficiency factor $Q$ -- the
fraction of photons averaged over the appropriate waveband that enter
the detector and go on to produce photoelectrons and so result in a
signal and $F$, the fraction of photons that survive the attenuation
due to Mie and Rayleigh scattering as the photons travel towards the
detector.  $Q$ varies among detectors. Here $Q$ is taken to be 0.2 and
$F$ is taken to be 0.5, conservatively.  The noise can be modeled as a
Poisson distribution with contributions from the source and the
background
\begin{equation}\label{SNRmain}
N_{noise} = (Q (F N_{\gamma} +  N_{background}))^{\frac{1}{2}}\,,
\end{equation}
where 
\begin{equation}\label{noise}
N_{background}  = \Phi_{background} A_{detector} \tau_{bin} \Omega_{pix}.
\end{equation} 
Here $\Phi_{background}$ is the number of background photons per unit area per unit time per unit solid angle
and $\Omega_{pix}$ is the solid angle subtended by a pixel.  
If  
\begin{equation}\label{lumfordet}
\frac{Q \ F N_{\gamma}}{N_{Noise}} \geq \mbox{SNR}_{min} 
\end{equation}
we can detect a macro. 
As is customary, we take $\mbox{SNR}_{min}=5$.

A combination of (\refeq{eventrate}), (\refeq{freestreamingevaluated}), 
(\refeq{bbevaluated}) and (\refeq{lumfordet}), 
applied to a particular FD telescope 
gives us the parameter space of macros that can be probed by that FD. 
In particular, we compare $N_{\gamma}$ with $N_{noise}$ and determine the range of $\sigma_x$  for which \eqref{lumfordet} holds.
We then determine the mass range that can be probed 
from the expected event rate in that FD.

\section{Specific examples}
We apply the above detection scheme to two specific examples,  
representing two main variants of fluorescence detection telescopes: 
the existing Auger Fluorescence Detector -- a ground-based fly's eye type instrument -- 
and the planned JEM-EUSO Fresnel-lens based space telescope.

\subsection{Pierre Auger Observatory}
The Pierre Auger Observatory~\cite{n1,n2} includes 24
Fluorescence Detector (FD) telescopes. Each telescope is composed of
22$\times$ 20 pixels covering a 30$\degree\times$ 30$\degree$ field of
view~\cite{n} out to a range of approximately 20 km.  Each telescope
thus observes a fiducial volume corresponding to a flat-sided pyramid
a square base with sides $2\times \sin{15\degree}\times$
20$\,$km$\approx 10^{4}\,$m.  We ignore the first km of the cone
nearest the telescope, as the time evolution of the intensity of the
ionization source as it moves through the atmosphere, will show no
significant variation from which to draw useful information.  We chose
a characteristic quantum efficiency $Q$ =20\% corresponding to an
average over the detected fluorescence wavelengths~\cite{x}.

The Auger detector was designed to look for relativistic cosmic ray particles.
The bin time was therefore set at $\tau_{bin}=100\,$ns, 
during which time an ultra relativistic cosmic ray would travel approximately $30\,$m, 
but a macro would traverse only a few cm.
For the much more slowly moving macros, we will need to increase this to
$\tau_{bin}=100\,\mu\,$s. 
This reconfiguration might require both software and hardware changes 
to the FD telescope and/or trigger system.

The analysis undertaken in section 2 was to 
determine the detectability of a passing macro in 
one pixel. To ensure that a passage of a macro 
will trigger at least 4 pixels in a row, we take 
a reduced detector volume by ignoring the outer 3 
pixels in our array. This reduces the detector 
volume in Auger by a factor of 
$\frac{17\times19}{20\times22} \approx 0.73$. 

Since we consider an isotropic flux we must account for all possible paths within this detection volume. The effective target area averaged over all angles for a single FD telescope is $120\,$km$^2$. Running the detector for an observation period of 1 year 
($\sim$ 10 years with the Auger FDs characteristic 
10$\%$ duty cycle\cite{n})
yields an expected number of events
\begin{equation} \label{mass}
	N_{events}^{1FD}= \left(\frac{g}{M_x}\right)\left(\frac{D_{max}(\sigma_x)}{km}\right)^2\,.
\end{equation}
where $D_{max}(\sigma_x)$ represents the maximum 
distance away we could detect a macro of a 
particular cross section. The minimum height 
below which no new parameter space can be probed 
is D$\approx 5.8\,$km. For heights below this, 
the effective target area of the detector is too 
small to extend above 55$\,$g, below which all 
values of $\sigma_x$ has been ruled out(see 
Figure \ref{fig:exclusion}) through CMB and mica 
measurements\cite{e}. The cost of probing higher 
masses, $M_x$, at higher values of D is a reduced 
sensitivity in $\sigma_x$ because of geometrical 
spreading. Thus, we choose D$=10$km in plotting 
Figure \ref{fig:sigmadependence} to illustrate 
the expected SNR for macros in unconstrained 
parameter space.

 We might be able to probe masses up to 660$\,$g with one Auger FD telescope. 
 If the entire Auger FD array could be used, 
 the effective target area becomes $3\times 10^3\,$km$^2$. 
 This pushes the upper mass accessible to $1.6 \times 10^4\,$g. 
 Another $50\,\%$ improvement could be obtained if the duty cycle could be improved to
$\sim15\,\%$, as has been claimed \cite{o}. 

R. Caruso et. al \cite{p} measured the background sky photon flux at two FD telescope sites at Malarg\"{u}e: Los Leones and Coihueco. 
Using \eqref{noise} and taking their highest measured background value of 
$\Phi_{background}= 134\,$m$^{-2}\,$deg$^{-2}\,\mu$s$^{-1}$, 
\begin{equation}\label{backAuger}
	\begin{aligned}
N_{background}^{1FD} 
&\approx 4\times 10^5
	\end{aligned}
\end{equation}
(for $A_{detector} \approx 13\,$m$^2$ \cite{q},
$\tau_{bin} = 100\,\mu$s and $\Omega_{pix} = (1.5\degree)^2$).

Using \eqref{freestreamingevaluated}, \eqref{SNR}, \eqref{SNRmain} and \eqref{backAuger} we find that 

\begin{figure}
 \centering

        \includegraphics[width=4in]{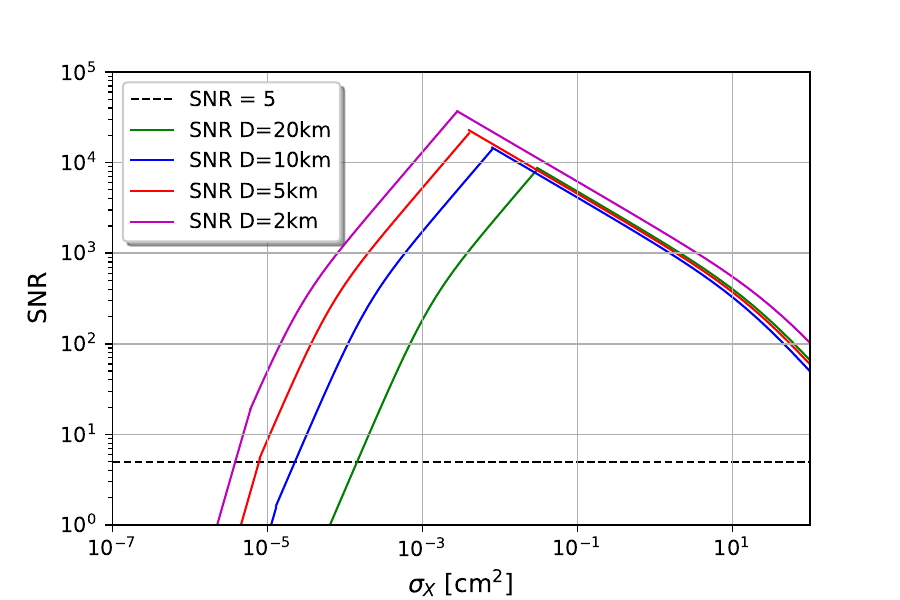}
        \caption{SNR as a function of $\sigma_x$ for various values of D(km) for Auger.}
        \label{fig:Auger}

 \end{figure}

\begin{subnumcases}
{{\hspace*{-0.5cm}\small SNR_{FS,Auger}^{1FD}}=}
\sqrt{\frac{3\times 10^{20} \left(\frac{km}{D}\right) \left(\frac{\sigma_x}{cm^2}\right)^{3}e^{-\frac{2D}{5\,km}}}{1+ 8 \times 10^{-16} \frac{D}{km}\left(\frac{cm^2}{\sigma_x}\right)^{3}e^{\frac{2D}{5\,km}}}}, &
$
\begin{split}
{\small \hspace*{-0.4cm} \text{\mbox{\bf FSU}}} 
\end{split}$        \\
\sqrt{\frac{2\times 10^{15} \left(\frac{km}{D}\right)\left(\frac{\sigma_x}{cm^2}\right)^{2}e^{-\frac{3D}{10\,km}}}
{1 + 2\times 10^{-10} \frac{D}{km}\left(\frac{cm^2}{\sigma_x}\right)^{2}e^{\frac{3D}{10\,km}}}}, &
$
\begin{split}
{\small \hspace*{-0.4cm} \text{\mbox{\bf FSUT}}}
\end{split}$    \\
\sqrt{\frac{2\times 10^{7} \left(\frac{km}{D}\right)\left(\frac{\sigma_x}{cm^{-1}}\right)^{2}e^{\frac{D}{10\,km}}}
{1 + 2\times 10^{-2} \frac{D}{km}\left(\frac{cm^2}{\sigma_x}\right)^{-1}e^{-\frac{D}{10\,km}}}}, &
$
\begin{split}
{\small \hspace*{-0.4cm} \text{\mbox{\bf FSST}}}
\end{split}$

\end{subnumcases}
where we have set $v_x = 250$kms$^{-1}$ to simplify the process of producing Figure \ref{fig:Auger}. We will discuss how reasonable is this assumption in section 5. Here we use the labels corresponding to the different emission
mechanisms and instrumental constraints: {\bf FS}=''Free-Streaming'',
{\bf FSUT}=''Free Streaming Unsaturated with $\tau_{bin} < t_{I0}$'' and {\bf
  FSST}=''Free Streaming Saturated with $\tau_{bin} < t_{I0}$''.  These ranges and expected
  sensitivities are also summarized in Table~1.

Using the expressions for the SNR, one can(for various values of D), by setting $SNR \geq 5$ for detection obtain the range of $\sigma_x$ that can be probed by Auger. 
We plot the SNR in Figure 
\ref{fig:Auger} for various values of D(km) for 
which detection is possible using the FDs of 
Auger. Although the expression for the SNR is labeled 
as applying only to 1 FD, it is also relevant to 
the entire array. Utilizing the entire array 
increases in the upper bound of masses that may 
be probed, which in turn increases the 
sensitivity to macro cross section because of a 
reduction in geometric spreading of the signal.




To determine the form the lower bound of the one-telescope purple region in Figure \ref{fig:exclusion},
we have iterated over various values $\sigma_x$ 
For each value of $\sigma_x$, we have iterated 
over various $v_x$ values and determined the 
maximum value of D(km) that a macro of the given 
$\sigma_x$ can be detected. For each $v_x$, the 
fraction of macros that would possess that speed 
is determined and this quantity multiplied to 
determine the through value of $M_x$ that can be 
probed by Auger. The highest $M_x$ for a given $\sigma_x$ was used in plotting Figure \ref{fig:exclusion}.

For the full Auger FD array, 
the lower boundary is of the same form that of the one FD bound, with the exception being that we can probe up to $1.2\times 10^4\,$g instead of $480\,$g. A similar analysis was done as in the one FD case.
The lower bound of the entire array region in Figure \ref{fig:exclusion} is the sum of the striped and non-striped purple regions. 

For the optically thick (black body) case using \eqref{bbevaluated}, \eqref{SNR}, \eqref{SNRmain} and \eqref{backAuger} we have for both a single FD telescope and the entire array



\begin{equation}
\begin{aligned}
SNR_{BB,Auger}^{1FD}=\sqrt{\frac{2\times 10^{12} \left(\frac{km}{D}\right) \left(\frac{\sigma_x}{cm^2}\right)^{-1}e^{\frac{D}{10\,km}}}{1+ 2 \times 10^{-7} \frac{D}{km}\left(\frac{cm^2}{\sigma_x}\right)^{-1}e^{-\frac{D}{10\,km}}}} 
\end{aligned}
\end{equation}



We have not shown the expressions for the transition between the optically thin and thick emission mechanisms because the interpolation must be done for specific values of D. 
For Auger, the BB emission mechanism lies beyond the range of parameter space that can be probed.
We plot the above expressions in Figure \ref{fig:Auger}.

In the following subsection, we repeat the calculations done for Auger above for JEM-EUSO.

\subsection{JEM-EUSO}
JEM-EUSO is a planned Ultra-High Energy Cosmic Ray 
	(UHECR, $E>10^{18}\,$eV) detector 
to be placed in the Japanese Experiment Module 
	of the International Space Station. 
It will watch the dark-side of the Earth and detect UV photons emitted in extensive air showers caused by  UHECRs, 
and especially Extremely High Energy Cosmic Ray (EHECR) 
	particle ($E\gtrapprox10^{20}$ eV). 
The telescope will have approximately $200,000$ pixels, 
a detection distance from space to Earth's surface of 400$\,$km, 
and an angular resolution of $0.07\degree$. 
This gives a fiducial cone with $L\simeq750\,$m at the ground. 
The planned JEM-EUSO binning time \cite{r} is 
$\tau_{bin}^{EUSO} = 2.5\,\mu$s.  As is the case for Auger, the JEM-EUSO binning time will
need to be increase by a factor of 1000 to a value of $2.5\,$ms for the macro search. We take D$=400$km because using a smaller D would reduce the strength of the signal as the signal depends strongly on the electron number density even though there is a reduced distance between the source and detector.

JEM-EUSO will also be able to operate in ``tilt'' mode, 
looking not straight down at the Earth but at an angle to the nadir. 
This will increase the effective area markedly. 
Though the consequence of this have yet to be fully explored \cite{s}, we will proceed with  $A_{ef} \sim 10^6 \,$km$^2$,
appropriate to a tilt angle of $40\degree$  
(see Figure 5 of \cite{t}). 
Equation \eqref{eventrate} gives, for one year of JEM-EUSO observations,
\begin{equation}
N_{events} \simeq \frac{6\times10^6 g}{M_x} \,,
\end{equation}
allowing us to probe masses up to $6 \times 10^6\,$g.

Meanwhile, based on \cite{u} and  \cite{v},
$\Phi_{background} = 600\,$m$^{-2}\,$sr$^{-1}\,$ns$^{-1}$,
$A_{detector} \approx 5\,$m$^2$ ,
$\tau_{bin} = 2.5\,$ms and $\Omega_{pix} = (0.07 \degree)^2$,
yielding
\begin{equation}\label{backJEM}
	\begin{aligned}
N_{background} & \approx 10^4.
	\end{aligned}
\end{equation}

Using \eqref{freestreamingevaluated}, \eqref{SNR}, \eqref{SNRmain} and \eqref{backJEM}, 
we find that 
\begin{subnumcases}
{{\hspace*{-0.5cm}\small SNR_{FS,JEM}^{1FD}}=}
\sqrt{\frac{3\times 10^{20} \left(\frac{km}{D}\right) \left(\frac{\sigma_x}{cm^2}\right)^{3}e^{-\frac{2D}{5\,km}}}{1+ 2 \times 10^{-17} \frac{D}{km}\left(\frac{cm^2}{\sigma_x}\right)^{3}e^{\frac{2D}{5\,km}}}}, &
$
\begin{split}
{\small \hspace*{-0.4cm} \text{\mbox{\bf FSU}}} 
\end{split}$        \\
\sqrt{\frac{2\times 10^{15} \left(\frac{km}{D}\right)\left(\frac{\sigma_x}{cm^2}\right)^{2}e^{-\frac{3D}{10\,km}}}
{1 + 4\times 10^{-12} \frac{D}{km}\left(\frac{cm^2}{\sigma_x}\right)^{2}e^{\frac{3D}{10\,km}}}}, &
$
\begin{split}
{\small \hspace*{-0.4cm} \text{\mbox{\bf FSUT}}}
\end{split}$    \\
\sqrt{\frac{2\times 10^{7} \left(\frac{km}{D}\right)\left(\frac{\sigma_x}{cm^{-1}}\right)^{2}e^{\frac{D}{10\,km}}}
{1 + 4\times 10^{-4} \frac{D}{km}\left(\frac{cm^2}{\sigma_x}\right)^{-1}e^{-\frac{D}{10\,km}}}}, &
$
\begin{split}
{\small \hspace*{-0.4cm} \text{\mbox{\bf FSST}}}
\end{split}$

\end{subnumcases}
To obtain the exclusion region in \ref{fig:exclusion}, $D=400$km was used. For various values of $v_x$, the minimum $\sigma_x$ that was detectable was obtained and the corresponding $M_x$ found using the fraction of macros having the minimum speed $v_x$.

\begin{figure}[h]
	\centering
	
	\includegraphics[width=4in]{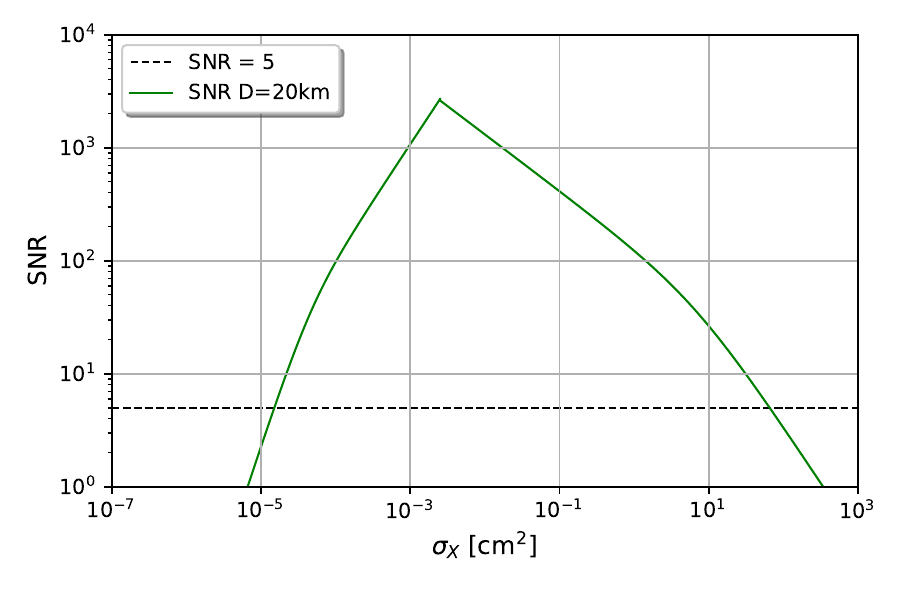}
	\caption{SNR as a function of $\sigma_x$ for $D=400\,$km for JEM-EUSO.}
	\label{fig:JEM}
	
\end{figure}

For the blackbody spectrum, we find that \eqref{bbevaluated}, \eqref{SNR}, \eqref{SNRmain} and \eqref{backJEM}  gives

\begin{equation}
\begin{aligned}
SNR_{BB,JEM}^{1FD}=\sqrt{\frac{2\times 10^{12} \left(\frac{km}{D}\right) \left(\frac{\sigma_x}{cm^2}\right)^{-1}e^{\frac{D}{10\,km}}}{1+ 4 \times 10^{-9} \frac{D}{km}\left(\frac{cm^2}{\sigma_x}\right)^{-1}e^{-\frac{D}{10\,km}}}}, 
\end{aligned}
\end{equation}

Again, we use the labels corresponding to the different emission
mechanisms and instrumental constraints: {\bf FS}=''Free-Streaming'',
{\bf FSUT}=''Free Streaming Unsaturated with $\tau_{bin} < t_{I0}$'' and {\bf
  FSST}=''Free Streaming Saturated with $\tau_{bin} < t_{I0}$''.  These ranges and expected
  sensitivities are also summarized in Table~2.

We plot $SNR_{BB,JEM}$ in Figure \ref{fig:JEM}.

In Figure \ref{fig:exclusion} below, we show how the regions of parameter space that could
potentially  be probed by Auger and JEM-EUSO fit into the existing
constraints on the macro parameters from  \cite{g}.  

It is of particular significance that both detectors are sensitive to macros
of nuclear or lower density, 
since the expected Standard Model macro candidates, 
as well as most others that have been explored 
(excepting primordial black holes),
are expected to be of approximately nuclear density
(see e.g. \cite{d}).

\section{Discussion}
In this section, we review some of the assumptions made in the preceding analysis.

We begin with the assumption that $L = D sin\theta$. $L$ is dependent on the inclination of the macro trajectory $i$, the polar angle $\phi$ and the height $D$. 

From geometrical considerations, we find
\begin{equation}
L = 2 D \frac{sin\left(\frac{\theta}{2}\right)}{cos(\frac{\theta}{2} - i)}(1 + cos\theta^2 tan(\phi)^2)^{\left(\frac{1}{2}\right)}
\end{equation}
A Monte Carlo simulation was done by sampling values of $\phi$ and $i$. The results were compared to the quantity $D sin\theta$. 
Approximately three quarters of the trajectories had a $1 \leq \frac{L}{D sin\theta} \leq 4$. With most of the trajectories being clustered around the quantity $D sin\theta$, we used $L = D sin \theta$ to simplify our calculations of the SNR.

This produces an underestimate of the true SNR because we are
approximating the entirety of photon production at distances greater
than the true production. For macro trajectories with $L \gg D
sin\theta$ the number of photons will vary significantly along the
trajectory. However, such trajectories make up a small fraction of all
trajectories (assuming an isotropic distribution of trajectories) so
that neglecting these will not detract significantly from our analysis
or the results.




 \begin{figure}
 \centering
 \begin{minipage}[c]{\textwidth}
 \centering
        \includegraphics[width=4.0in]{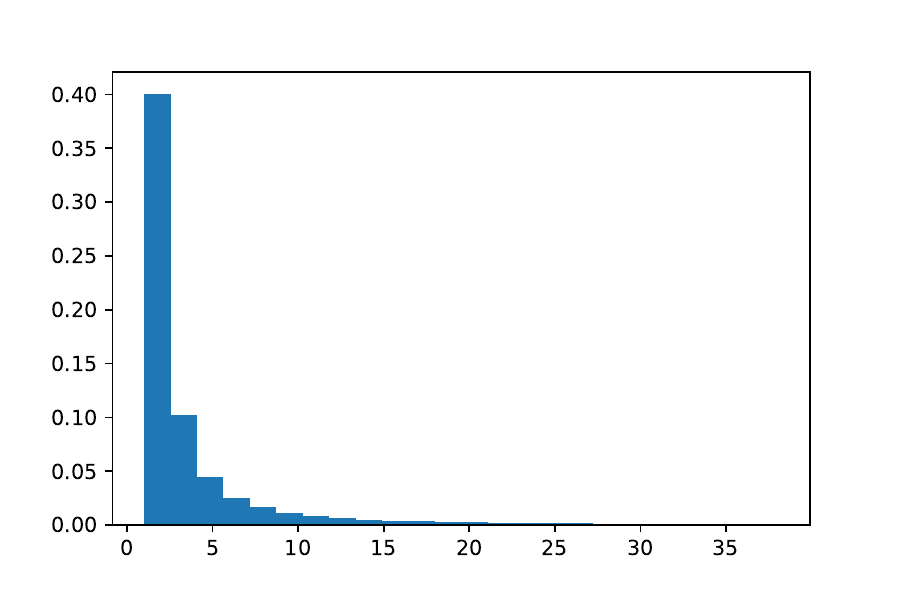}
        \caption{Monte Carlo results comparing the ratio of the actual path length seen by a pixel to the approximation $L = Dsin\theta$.}
        \label{fig:histogram}
 \end{minipage}
 \end{figure}

\label{sec:fofvx}
The next assumption involved the presentation of results by
approximating $v_x = 250\,$km s$^{-1}$.  In fact, as discussed in the
introduction, we expect that dark matter macros will have a Maxwellian
velocity distribution as described
by~\eqref{eq:maxwellian}. Furthermore, we expect this distribution
to be somewhat modified due to the Earths motion; specifically the
Earth is moving around the Sun at $\sim 30\,$km s$^{-1}$ and that the
Sun (and hence the Solar System) is moving around the center of the
Milky Way at $\sim 220\,$km s$^{-1}$.  As a result, we expect that
there should be no significant contribution to the distribution for
$v_x \lesssim 170\,$km s$^{-1}$.  To determine the impact of
these effects on the velocity distribution as detected at the Earth,
we convolve of the macro Maxwellian distribution and the Earth's speed
distribution.\footnote{Specifically, this convolution gives a modified
  velocity distribution as follows:

\begin{subnumcases}
{{\hspace*{-0.5cm}\small f(v_x)}=}
\left(31250\ {\tiny \frac{\mbox{km}^2}{\mbox{s}^2}}\right) \times  \left\{(-170\ km/s + v_x) \exp \left[-\left(\frac{-170\ km/s + \nonumber v_x}{250\ km/s}\right)^2\right] \right.  + \\
   \left. (125\ km/s) \sqrt{\pi} \Erf\left(\frac{-170\ km/s + v_x}{250\ km/s}\right)\right\}, &
   
$
\begin{split}
  {\small \hspace*{-3.4cm} \text{\mbox{$170 \leq v_x \leq 230\,$km s$^{-1}$}}} \\ 
\end{split}$   \\
\left(31250\ {\tiny \frac{\mbox{km}^2}{\mbox{s}^2}}\right) \times \left\{(-230\ km/s + v_x)\exp \left[-\left(\frac{-170\ km/s + \nonumber v_x}{250\ km/s}\right)^2\right] \right. - \\ \nonumber
  (-170 km/s + v_x)\exp\left[-\left(\frac{-170\ km/s + v_x}{250\ km/s}\right)^2\right]  - \\ \nonumber
  (125\ km/s) \sqrt{\pi} \Erf\left(\frac{-230\ km/s + v_x}{250\ km/s}\right) + \\
  \left.(125\ km/s) \sqrt{\pi} \Erf\left(\frac{-170\ km/s + v_x}{250\ km/s}\right)\right\}, &
$
\begin{split}
{\small \hspace*{-3.4cm} \text{\mbox{\mbox{$v_x \geq 230\,$km s$^{-1}$}}}}
\end{split}$    
\end{subnumcases}
} This yields a velocity distribution where the majority of dark matter macros should have a velocity range between 
$170$~km s$^{-1}$ and $350$~km s$^{-1}$.

 \begin{figure}
 \centering
 \begin{minipage}[c]{\textwidth}
 \centering
        \includegraphics[width=4.0in]{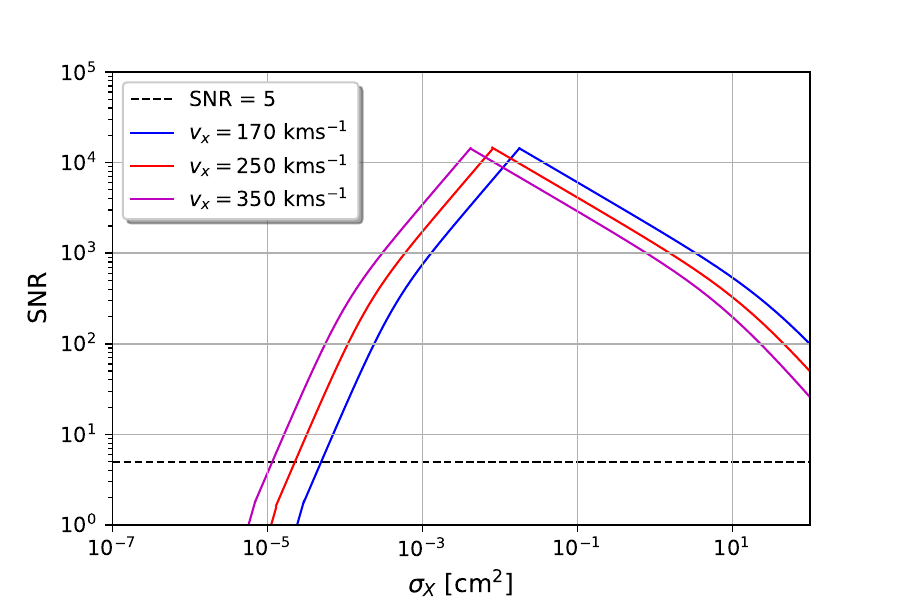}
        \caption{The SNR vs $\sigma_x$ for $D=10\,$km and varying $v_x$}
        \label{fig:vx}
 \end{minipage}
 \end{figure}

Figures \ref{fig:sigmadependence}, \ref{fig:Auger} and \ref{fig:JEM}
were generated assuming $v_x = 250\,$km s$^{-1}$.  In
Figure~\ref{fig:vx} we reproduce Figure \ref{fig:Auger} with a fixed
$D=10\,$km, but varying $v_x$.  Here we see that the results presented
in the preceding sections are a reliable measure for a wide range of
realistic $v_x$ values. Only for small values of $v_x$, which make up
a small fraction of the trajectories would there be significant
deviation from the graphical results presented before this.

If a macro were to be detected, $v_x$ would be 
inferred from the motion of the macro and this 
would then be used to constrain the range of 
$\sigma_x$.

We have also taken into account the effects of attenuation of the
signal as the photons propagate from the point of production to the
detector. However, these effects have do not have a major impact on
the our results.  Attenuation due to Rayleigh scattering is easily
calculated as a function of viewing angle and distance and corresponds
to a maximum attenuation of about 70\% of signal at the edge of our
20~km distance maximum fiducial distance for the FD detectors of
Auger~\cite{x}.  A typical value of Rayleigh attenuation corresponding
to 0.5 is quite conservative and applied for both Auger and JEM-EUSO.
Mie scattering due to ground-hugging aerosols can also contribute to
attenuation.  Atmospheric monitoring data taken over nine years of
data using the Auger Central Laser Facility (CLF) show a typical
aerosol optical depth is 0.04 even out to a vertical equivalent
atmosphere of 10~km \cite{y}.  For Auger at 20~km this corresponds to
an additional attenuation of
$\left[1-\exp^{-0.04*\left(\frac{20}{10}\right)}\right]=$ 8\%.  Even
on rare nights when the scattering optical depth exceeds 0.1,
additional attenuation will not exceed 20 percent of our signal.  The
effect of aerosol scattering is even less prominent for JEM-EUSO which
is viewing showers at a near-vertical angle from space.  \color{black}

\section{Conclusion}
     
Macroscopic dark matter is a broad class of 
alternatives to particulate dark matter that, 
compellingly, includes plausible Standard Model 
candidates. 
The passage of a macro  through Earth's atmosphere will cause dissociation
and ionization of air molecules, resulting, through recombination, 
in a signal visible to Fluorescence Detectors such as those used to search for Ultra High Energy Cosmic Rays. As for such UHECR, large effective target areas are 
necessary to compensate for the low maximum flux of macros. Unlike UHECR,
macros would be expected to travel several times faster than typical 
solar system objects, such as meteoroids, but still very non-relativistically.
Existing and planned cosmic ray detectors would therefore need to make software,
or possibly hardware, accommodations in order to detect the more slowly traced-out
trajectories of macros. If they do, they have significant discovery potential
for macroscopic dark matter of nuclear or greater density, 
including the most compelling non-black-hole candidates,
able to probe up to masses of several tonnes, compared to current lower
limits of just several tens of grams.

\acknowledgments GDS and JSS thank David Jacobs for initial
considerations about this project.  CEC thanks the members of the
Pierre Auger publication committee, especially Roger Clay, for helpful
review suggestions.  This work was partially supported by Department
of Energy grant DE-SC0009946 to the particle astrophysics theory group
at CWRU.

\clearpage

\begin{table}
\centering
\begin{minipage}[c]{\textwidth}
\centering
\begin{tabular}{||l|l||c|c||} \hline
  \multicolumn{2}{|c||}{Emission}      & \multicolumn{2}{c||}{Pierre Auger Observatory}  \\
   \multicolumn{2}{|c||}{Mode}         & \multicolumn{2}{c||}{with $\tau_{bin}=100\mu s$}   \\ \hline
    &                       & Cross-Section (cm$^2$)                                  & Sensitivity      \\ \hline
FS  &  Free-Streaming       & $10^{-6} \leq \sigma_x \leq 10^{-5}$                    &   Good           \\
    &  (optically thin)     &                                                         & SNR $\geq 5$     \\
    &                       &                                                         &  $10^{-2}  \lesssim M_x  \lesssim 10^{4}$ g  \\ \hline
FSUT & Free-Streaming    & $10^{-5} \leq \sigma_x \leq 10^{-2}$             &  Strong          \\
    &   Unsaturated          &                                                         &  SNR $>> 5$      \\ 
    & $\tau_{bin} \leq t_{I0}$           &                                                         &  $10^{-2}  \lesssim M_x  \lesssim  10^{4}$ g  \\ \hline
FSST &  Free-Streaming          & $ 10^{-2} \leq \sigma_x \leq  10^{1}$  &  Strong          \\
    &  Saturated    &                                                         &   SNR $>> 5$     \\
    & $\tau_{bin} \leq t_{I0}$           &                                                         & $1 \lesssim M_x  \lesssim 10^{4}$ g  \\ 
 \hline
\end{tabular}
\caption{Tabular summary of sensitivity for Auger for different emission mechanisms assumed over a range of
  cross-section values. Rows correspond to ranges of cross-section and SNR calculations as delineated in Equations~(4.3),~(4.7a),~(4.7b),~and~(4.7c).} 
\end{minipage}
\end{table}

\begin{table}
\centering
\begin{minipage}[c]{\textwidth}
\centering
\begin{tabular}{||l|l||c|c||} \hline
  \multicolumn{2}{|c||}{Emission}      & \multicolumn{2}{c||}{JEM-EUSO}  \\
   \multicolumn{2}{|c||}{Mode}         & \multicolumn{2}{c||}{with $\tau_{bin}=2.5 ms$}   \\ \hline
    &                       & Cross-Section (cm$^2$)                                  & Sensitivity      \\ \hline
FS  &  Free-Streaming       & $10^{-6} \leq \sigma_x \leq 10^{-5}$                    &   Good   \\
    &  (optically thin)     &                                                         &   SNR $\sim$ 5                 \\
    &                       &                                                         &  $ M_x \leq 10^6$ g                 \\ \hline
FSUT & Free-Streaming    & $10^{-5} \leq \sigma_x \leq 10^{-2}$             &  Good         \\
    &   Unsaturated          &                                                         &  SNR  $\geq 5$      \\ 
    & $\tau_{bin} \leq t_{I0} $           &                                                         &  $M_x  \lesssim  10^{7}$ g  \\ \hline
FSST &  Free-Streaming          & $ 10^{-2} \leq \sigma_x \leq 10^{2}$  &  Strong          \\
    &  Saturated    &                                                         &   SNR $>> 5$     \\
    &  $\tau_{bin} \leq t_{I0} $     &                                                         & $M_x  \lesssim 10^{7}$ g  \\ \hline
\end{tabular}
\caption{Tabular summary of sensitivity for JEM-EUSO for different emission mechanisms assumed over a range of
  cross-section values. Rows correspond to ranges of cross-section and SNR calculations as delineated in Equations~(4.10),~(4.11a),~(4.11b),~(4.11c),~and~(4.11d).}
\end{minipage}
\end{table}

\clearpage

 \begin{figure}
 \centering
 \begin{minipage}[c]{\textwidth}
 \centering
        \includegraphics[width=6.0in]{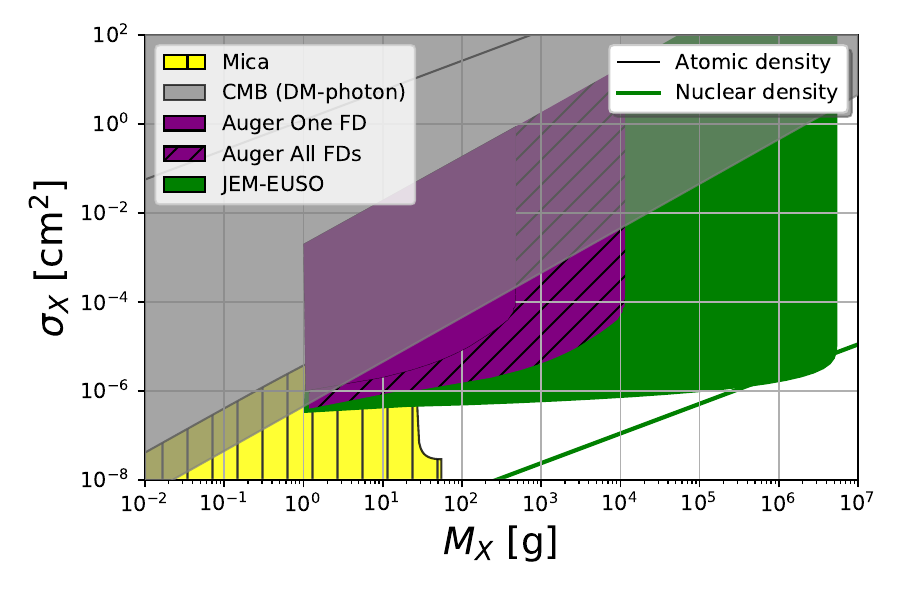}
        \caption{Figure 3 of \cite{g} with the various regions of parameter space that could be probed by both PA(for one FD telescope in purple and the full array in purple with diagonal lines) and JEM-EUSO (in green). }
        \label{fig:exclusion}
 \end{minipage}
 \end{figure}

\clearpage


\begin{thebibliography}{99}
	
	\bibitem{a}
	Lelli, Federico and McGaugh, Stacy S. and Schombert,
	James M, \emph{SPARC: Mass Models for 175 Disk Galaxies with Spitzer
		Photometry and Accurate Rotation Curves,}, arxiv:1606.09251
	
	\bibitem{b}
	Witten E, \emph{Cosmic Separation of Phases,}, \emph{Phys. Rev. D} {\bf 30:}{272-285} (1984)
	
	\bibitem{c}
	Bryan W. Lynn and Ann E. Nelson and Nikolaos Tetradis, \emph{Strange Baryon Matter,} \emph{Nuc. Phys. B} {\bf 345:}{186-209} (1990)

	\bibitem{d}
	Ariel R. Zhitnitsky, \emph{"Nonbaryonic" Dark Matter as Baryonic Color Superconductor}, \emph{JCAP} 
	arxiv:hep-ph/0202161
	
	\bibitem{e}
	David M. Jacobs and Glenn D. Starkman and Bryan W. Lynn, \emph{Macro Dark Matter}, \emph{MNRAS} {\bf 450:}{3418–3430} (2015) 
	arxiv:1410.2236


	\bibitem{f}
	David M. Jacobs and Glenn D. Starkman and Amanda Weltman, \emph{Resonant bar detector constraints on macro dark matter}, \emph{Phys. Rev. D} {\bf 91:}{115023} (2015)
	arxiv:1504.02779
	
	\bibitem{g}
	David Cyncynates and Joshua Chiel and Jagjit Sidhu and Glenn D. Starkman, \emph{Reconsidering seismological constraints on the available parameter space of macroscopic dark matter}, \emph{Phys. Rev. D} {\bf 95:}{063006} (2017) 
	arxiv:1610.09680
	
	\bibitem{h}	
	Capitelli, M and Colonna, G and Gorse, C and D'Angola, A \emph{Transport properties of high temperature air in local thermodynamic equilibrium}, \emph{Euro. Phys. Jour. D} {\bf 11:}{279-289} {2000}
	
	\bibitem{i} 
	Eisazadeh-Far, Kian and Metghalchi, Hameed and Keck, James C. \emph{Thermodynamic Properties of Ionized Gases at High Temperatures}
	\emph{Jour. of Ener. Res. Tech.} {\bf 133:}{022201} {2011}
	
	\bibitem{j}
	Armstrong, B H and Johnston, R R and Kelly, P S, \emph{Opacity of High Temperature Air}
	Lockheed Missiles And Space Company {1965}
	
	\bibitem{j1}
	A. Bourdon and P. Vervisch, \emph{Three-body recombination rate of atomic nitrogen in low-pressure plasma flows}, \emph{American Physical Society}, {\bf 54:}{1888-1898} 1996
	


	
	
	\bibitem{m}
	A. Kramida and Yu. Ralchenko and J. Reader and NIST ASD Team, \emph{NIST Atomic Spectra Database (ver. 5.3) , [Online]. Available:
		{\tt{http://physics.nist.gov/asd}} [2017, July 21].
		National Institute of Standards and Technology,
		Gaithersburg, MD.}, {2015}
	
        \bibitem{n1}  The Pierre Auger Collaboration, \emph{The Pierre Auger Observatory},  \emph{Nucl. Instrum. Meth. A} {\bf 798:}{172-213} (2015) arxiv:1502.01323

        \bibitem{n2}  The Pierre Auger Collaboration, \emph{The Pierre Auger Observatory Upgrade - Preliminary Design Report},  (2016) arxiv:1604.03637

        \bibitem{n}
        Argiro, Stefano \emph{Performance of the Pierre Auger Fluorescence Detector and Analysis of Well Reconstructed Events}, \emph{Universal Academy Press, Inc} {p. 457-460} (2003)


	\bibitem{o}
	Hermann-Josef, T \emph{The HEAT Telescopes of the Pierre Auger Observatory Status and First Data,} \emph{32nd International Cosmic Ray Conference Beijing} {2011}
	
	\bibitem{p}
	R. Caruso and A. Insolia and S. Petrera and P. Privitera and F. Salamida and V. Verzi \emph{Measurement of the Sky Photon Background Flux at the Auger Observatory}, \emph{29th International Cosmic Ray Conference Pune} {2005}
	
	\bibitem{q}
	The Pierre Auger Collaboration, \emph{The Fluorescence Detector of the Pierre Auger Collaboration,} \emph{Nucl. Instrum. Meth. A} {\bf 620:}{227-251} {2010}
	arxiv:0907.4282
		
	\bibitem{r}
	Y. Takahashi and JEM-EUSO Collaboration \emph{The Jem-Euso Mission,} \emph{New J. Phys} {\bf 11:}{065009} (2009)
	arxiv:0910.4187

	\bibitem{s}
	A. Haungs and JEM-EUSO Collaboration \emph{Physics Goals and Status of JEM-EUSO and its Test Experiments,} \emph{J. Phys.: Conf. Ser.} {\bf 632:}{012092} (2015)
	arxiv:1504.02593
	
	\bibitem{t}
	JEM-EUSO Collaboration \emph{JEM-EUSO Observational Technique and Exposure,} \emph{SpringerLink} {\bf 40:}{117-134} (2015)
	
	\bibitem{u}
	O. Catalano et al. \emph{The atmospheric nightglow in the 300-400 nm wavelength: Results by the balloon-borne experiment} \emph{Nuclear Instruments and Methods in Physics Research Section A: Accelerators, Spectrometers, Detectors and Associated Equipment,} {\bf 480:}{547–554} (2002)
	
	\bibitem{v}
	Marco Ricci and JEM-EUSO Collaboration \emph{The JEM-EUSO Program,} \emph{J. Phys.: Conf. Ser.} {\bf 718:}{052034} (2016)

	\bibitem{w}
	H. Prieto-Alfonso and L. del Peral and M. Casolino and K. Tsuno and T. Ebisuzaki and M. D. Rodríguez Frías and JEM-EUSO Collaboration \emph{Multi Anode Photomultiplier Tube Reliability Assessment for the JEM-EUSO Space Mission,} \emph{Reliability Engineering and System Safety} {\bf 133:}{137-145} (2015)

	 
      \bibitem{x}
        The Pierre Auger Collaboration, {\em The Piere Auger Observatory}, \emph{Nucl. Instrum. Meth. A} {\bf 798:}{172-213} {2015}
        arxiv:1502.01323 (2015)

	\bibitem{y}
    V. Laura for the Pierre Auger Collaboration, 
    \emph{Atmospheric Aerosol Attenuation Measurements at the Pierre Auger Observatory,}
    \emph{International Workshop on Atmospheric Monitoring for High-Energy Astroparticle Detectors (AtmoHEAD 2013) Gif-sur-Yvette, France, June 10-12, 2013,}
    arxiv:1402.6186 (2014)

	
	
	
	
	
\end{thebibliography}
\end{document}